\documentstyle[times,graphics,epsfig,amsmath,amssymb,astrobib]{mn2e}

\newcommand{\be}{\begin{equation}}
\newcommand{\e}{\end{equation}}
\newcommand{\th}{{\bf \theta}}

\newcommand{\E}{\hat E}
\newcommand{\bear}{\begin{eqnarray}}
\newcommand{\ear}{\end{eqnarray}}
\newcommand{\nline}{\nonumber \\}
\newcommand{\f}{\frac}

\def\u{{\bf U}}
\def\apj{ApJ}
\def\aj{AJ}

\def\apjs{ApJS}
\def\mnras{MNRAS}

\def\u{{\vec U}}
\def\th{\vec{\theta}}

\def\E{\hat{E}}

\begin{document}
  \title[The optimal redshift for detecting  ionized bubbles]
{The optimal redshift for detecting  ionized bubbles in HI 21-cm maps}
\author[Datta, Bharadwaj \& Choudhury]
{Kanan K. Datta$^1$\thanks{E-mail: kanan@phy.iitkgp.ernet.in}, 
Somnath Bharadwaj$^1$\thanks{E-mail: somnathb@iitkgp.ac.in} and
T. Roy Choudhury$^2$\thanks{E-mail: tirth@hri.res.in}\\
$^1$Department of Physics and Meteorology \&  
Centre for Theoretical Studies, IIT, Kharagpur 721302, India\\
$^2$Harish-Chandra Research Institute, Chhatnag Road, Jhusi, Allahabad
211019, India}
\maketitle
\date{\today}
\begin{abstract}
The detection of individual ionized bubbles  in HI 21-cm maps is one
of the most promising, direct probes of  the epoch of reionization
(EoR). At least  $1000 \ {\rm hrs}$ of observation would be required 
for such a detection with  either the  currently functioning GMRT or 
the upcoming MWA. Considering  the large investment of telescope time  
it is essential to identify the``optimal redshift'' where the
prospects of a detection are most favourable. We find that the optimal 
redshift is determined by a combination of
instrument dependent factors and the evolution of the neutral fraction
$x_{\rm HI}$. 
We find that the redshift range $8.1 \pm 1.1$ and $9.8 \pm 1$ are
optimum for detecting ionized bubbles with the GMRT and MWA
respectively. The prospects of a detection, we find,  are more
favourable in a scenario with late reionization   with $x_{\rm HI}
\approx 0.5$ at $z \approx 7.5$ as compared to an early reionization
model where $x_{\rm HI} \approx 0.5$ at $z \approx 10$. 
In the late reionization scenario, for both instruments  a 
 $3 \sigma$ detection is possible for   bubbles of comoving radius $R_b
\ge 30 \ {\rm    Mpc}$  with $1000 \ {\rm hrs}$ of observation. 
 Future observations will either lead to the detection of ionized
 bubbles, or in the event of  non-detection, 
lead to constraints  on the product $x_{\rm HI} \ R_b^{\gamma}$ for
the observational volume, where $\gamma=1.5$ and $2$ for GMRT and MWA
respectively.
\end{abstract}
\begin{keywords}
cosmology: theory, cosmology: diffuse radiation, Methods: data analysis
\end{keywords}

\section{Introduction}
It is currently accepted  that the Universe was reionized by the
growth of  
ionized bubbles around  luminous sources in the redshift range $z\sim 6-15$ 
 \cite{fan06,choudhury06,komatsu08}.
Detection of individual ionized bubbles (HII regions) in HI 21-cm maps
of reionization is one of the major, important approaches that   
will be adopted
by the present and upcoming radio experiments (GMRT, MWA, LOFAR, SKA) to
probe the EoR. 
Such observations  will directly probe the properties of the ionizing
sources and the evolution of the surrounding  IGM
\cite{wyithe04a,wyithe05,maselli07,geil07}  and are expected
to complement the  study
of reionization through the power spectrum  of HI  brightness temperature
fluctuations. 
Detection of individual bubbles  is a big challenge
because the HI signal will be buried in strong foregrounds and system noise
\cite{ali08}. 

In an earlier paper \cite{datta2}, hereafter referred to as Paper
I, we have proposed a visibility based  matched filter technique to 
optimally combine the entire HI signal from an ionized  bubble while
removing  the foregrounds and minimizing system noise.  
Using visibilities  has an advantage  over image
based techniques because the system noise contribution in different
visibilities is independent whereas   the noise in different pixels of
a radio-interferometric images is not. Our investigations show that
for both the GMRT and the MWA, at redshift $z=8.5$,  it will be possible
to detect ionized bubbles  of comoving radius $R_b > 40 \, {\rm Mpc}$
and $R_b > 22 \, {\rm Mpc}$  in $100$ and $1,000$ hours of
observations respectively. We also find that 
fluctuations in the HI outside the bubble that we are trying to detect
impose  a fundamental restrictions on the smallest bubble that can be
detected. Assuming that the HI outside the bubble traces the dark
matter, we find that it will not be possible to detect bubbles with
comoving radius less than $8$ and $16 \, {\rm Mpc}$ with the GMRT and
the MWA respectively,  however large be the integration time.  In a
subsequent paper \cite{datta3}, hereafter referred to as Paper II) 
we have used simulations to validate our matched filter technique and
assess the impact of patchy reionization outside the bubble that we are
trying to detect on the bubble detection. 

The question ``What is the optimal redshift for bubble detection?'' is
particularly important when planning future observations. Estimates
show (Papers I and II) that at least 1000 hrs of observation will be
required for a detection with either the GMRT or the MWA. Considering
the large investment in observing time,  it is important to target the
redshift where the  prospect of a detection is most favourable. 
In addition, it
is important to have a clear picture of the different  factors that
contribute towards deciding  the most optimal  redshift. We expect
this  to provide insights  useful for  the design of future
observational programmes and also the  design of future  low-frequency
radio telescopes. 

A variety of redshift dependent factors influence the signal from an
ionized bubble. While a number of these pertain to the instrument in
question, the frequency dependence of the sky temperature and the
redshift evolution of the neutral hydrogen fraction also play an
important role. 
 In this paper we analyze all of the effects that
determine the optimal redshift for detecting ionized bubbles.  
We consider two different models for the redshift evolution of 
the neutral fraction and make predictions for the GMRT and the MWA. 

The paper is organized as follows. In Section 2 we briefly review the
matched filter technique for bubble detection.
In Section 3 we establish scaling relations for the matched filter
signal to noise ratio  (SNR) assuming an uniform  
baseline distribution. We also discuss the models of HI evolution that
we adopt.  We present our results and conclusions in Section 4.

Throughout out this paper we adopt cosmological parameters from
\cite{dunkley09}. For the GMRT we use  the antenna specifications
from their web site and for the MWA we use 
the instrumental parameters from \cite{bowman07}. 

\vspace{-0.5cm}
\section{The matched filter technique for detecting ionized bubbles
  in  redshifted  21-cm maps}   
The visibility  recorded in a radio-interferometric
observation of an ionized bubble can be written as
\be
V(\u,\nu)=S(\u,\nu)+HF(\u,\nu)+N(\u,\nu)+F(\u,\nu) \,.
\label{eq:vis2}
\e
Here we refer to $\u={\bf d}/\lambda$ as a baseline,  ${\bf d}$ being
the  physical separation between a pair of antennas projected on the
plane perpendicular to the line of sight and $\lambda$ is wavelength
corresponding to the observed frequency $\nu$. In eq. (\ref{eq:vis2})
$S(\vec{U},\nu)$ is the HI signal from the ionized  bubble, $
HF(\u,\nu)$ is the contribution from fluctuations 
in the  HI outside the target  bubble, $N(\vec{U},\nu) $ is the system
noise  and $F(\vec{U},\nu)$ is
the contribution from other astrophysical foregrounds. The contributions 
$HF(\u,\nu),N(\u,\nu)$ and $F(\u,\nu)$
are all assumed to be random variables with zero mean, whereby
$\langle V(\u,\nu) 
\rangle = \langle S(\u,\nu) \rangle$.  The angular brackets here 
denote average with respect to different realizations of the HI
fluctuations, system noise and foregrounds.

We consider a spherical ionized bubble of comoving radius $R_b$
centered at redshift $z_c$ located at the center of the field of view 
(FoV). The bubble is assumed to be embedded in an uniform IGM  with 
neutral hydrogen fraction $x_{\rm
  HI}$.  A bubble  of comoving radius $R_b$ will be seen as a
circular disc in  each of the
frequency channels that cut through the bubble. At a frequency channel
$\nu$, the  angular radius of the disc is 
$\theta_{\nu}=(R_b/r_{\nu}) \sqrt{1- (\Delta \nu/\Delta
  \nu_b)^2}$ where  
$\Delta \nu=\nu_c-\nu$ is the distance from the bubble's center $\nu_c
=1420 \, {\rm MHz}/(1+z_c)$ and 
$\Delta \nu_b=R_b/r^{'}_{\nu}$ is the bubble's radius  in  frequency
space. Here $r_{\nu}$ is the comoving distance corresponding to
$z=(1420 \, {\rm MHz}/\nu)-1$, and $r^{'}_{\nu}= dr_{\nu}/d\nu $. The
expected visibility signal $S(\u,\nu)$ in each frequency channel is
the Fourier transform of a circular disc  which can be
expressed in  terms of  $J_1(2\pi U \theta_{\nu})$ , the first
order Bessel function (Paper I). 
 In each channel,  the signal has a peak value $\mid S(0,\nu)\mid =\pi 
x_{\rm HI} \bar{I}_{\nu} \theta^2_{\nu} $ where  
 $\bar{I_{\nu}}=2.5\times10^2\frac{Jy}{sr} \left (\frac{\Omega_b 
  h^2}{0.02}\right )\left( \frac{0.7}{h} \right ) \left
(\frac{H_0}{H(z)} \right ) $ is the background HI specific intensity expected 
from completely  neutral medium. 
  The signal is  largely contained within 
baselines $U \le U_0   =0.61/\theta_{\nu}$ where  the
Bessel function  has its first zero   crossing, and  the signal is  much
smaller 
at larger baselines. The signal
$S(\u,\nu)$ picks up an extra phase if the bubble is shifted from the
center of the FoV. The  amplitude of the signal also   falls because
of the  telescope's  primary  beam pattern (Paper I), and in this
paper we restrict our analysis to the most favourable situation where
the bubble is at the center of the FoV.   
 The terms  $x_{\rm
  HI}$, $\bar{I_{\nu}}$, $\theta_\nu$ and  $\Delta \nu_b$  are all
redshift dependent, and hence  the signal too is strongly 
redshift dependent. In Section 3 we will discuss the combined effect
of all these factors on  bubble detection.

In order to detect an ionized bubble whose expected signal is $S(\u,\nu)$
we use the matched filter $S_f(\u,\nu)$ defined as 
\begin{eqnarray}
S_f(\u,\nu) \!\!\!\!\!&=& \!\!\!\!\! \left(\frac{\nu}{\nu_c}\right)^2 \left[ S(\u,\nu) -\right.\nonumber \\
&&\!\!\!\!\! \left. 
\Theta\left(1-2 \frac{\mid \nu - \nu_c \mid}{B'}\right)
\frac{1}{B'} \int_{\nu_c - 
  B'/2}^{\nu_c + B'/2} S(\u,\nu') \, d \nu' \right]. \nonumber \\
\end{eqnarray}
Note that the filter is constructed using the signal that we are 
trying to detect. The term $(\nu/\nu_c)^2$ accounts the frequency dependent
$U$ distribution for a given array.  The function $\Theta$ is the
Heaviside step function. The second term in the square brackets serves to
remove the foregrounds within the frequency range $\nu_c- B'/2$ to
$\nu_c+B'/2$.  Here $B'=4\,\Delta \nu_b$ is the frequency width that
we use to estimate and subtract out a frequency independent
foreground contribution. This, we have seen in Paper I, is adequate to
remove the foregrounds such that the residuals are considerably smaller than
the signal. Further we have assumed that $B'$ is smaller than the total
observational bandwidth $B$. The filter  
$S_f(\u,\nu)$  depends on $[R_b,z_c,\th_c]$ the  comoving  radius, 
redshift and angular position of the target bubble that we are trying
to detect. 

Bubble detection is carried out by combining the entire  observed 
visibility signal weighed with the filter.  The estimator  $\hat{E}$
is defined as  
\be
\hat{E}=  \left[ \sum_{a,b} S_{f}^{\ast}(\u_a,\nu_b)
\hat{V}(\u_a,\nu_b) \right]/\left[   \sum_{a,b} 1 \right] \,,
\label{eq:estim1}
\e
where the sum is over all frequency channels and baselines.  The 
expectation value $\langle \E \rangle$ is non-zero only if an ionized
bubble is present, and it is zero if there is no bubble  in the 
FoV. 

The  system noise (NS), HI fluctuations (HF) and the foregrounds (FG)
all contribute to the variance of the estimator 
\begin{eqnarray}
\langle (\Delta \E)^2 \rangle =\left <(\Delta \hat
E)^2 \right >_{{\rm NS}}+\left<(\Delta \hat E)^2 \right >_{{\rm HF
}} \,
 +\left<(\Delta \hat E)^2 \right >_{{\rm FG}} 
\label{eq:16}
\end{eqnarray}
A $3 \sigma$  detection is possible only if 
$\langle \hat{E} \rangle > 3 \sqrt{ \langle (\Delta \E)^2
\rangle}$.  In a situation  where this condition is satisfied,  the
observed value  $E_0$ may be interpreted as a  detection if $E_0
> 3 \sqrt{\langle (\Delta \E)^2   \rangle}$.

Because of our choice of the matched filter, 
the contribution from the residuals after foreground subtraction
$\left<(\Delta \hat E)^2 \right >_{{\rm  FG}}$ is predicted 
to be smaller than the signal (Paper I) and we 
do not consider it in the subsequent analysis.
The contribution $\left<(\Delta \hat E)^2 \right >_{{\rm HF
}}$ which arises from the HI fluctuations outside the target bubble
imposes a fundamental restriction on bubble detection. It is not
possible to detect an ionized bubbles for which  
$\langle \hat E  \rangle  \lesssim \sqrt{\left<(\Delta \hat E)^2
  \right >_{{\rm HF }}}$. Bubble detection is meaningful only in
situations where the 
contribution from HI fluctuations is considerably smaller  than the
expected  signal.  Once this condition is satisfied, it is the 
${\rm SNR}$ defined as 
\begin{equation}
{\rm SNR}=\langle \E \rangle/ \sqrt{\langle (\Delta \hat
E)^2 \rangle_{\rm NS}} 
\label{eq:snr}
\end{equation}
which is important for bubble detection. 
The value of SNR peaks when the parameters of the filter exactly match the
bubble that is actually present in the observation, and decreases
from its peak value if
there is a mis-match (Paper II). In the subsequent analysis we
shall use this to assess the redshift that is optimal for
bubble detection.

It is possible to analytically estimate 
$\langle \E \rangle$, $\langle (\Delta \hat E)^2 \rangle_{\rm NS}$ and
$\langle (\Delta \hat E)^2 \rangle_{\rm HF}$ in the continuum limit
(Paper I). We have 

\be
\langle \E \rangle  =\int d^2U \, \int d\nu  \, \rho_N(\u,\nu) \, \, 
{S_f}^{\ast}(\u,\nu) S(\u,\nu)  \,,
\label{eq:estim2}
\e  
\be
\langle (\Delta \hat E)^2 \rangle_{\rm NS}
= \sigma^2 
\int d^2U \, \int d\nu  \, \rho_N(\u,\nu) \, \, 
\mid S_{f}(\u,\nu)\mid^2\,.
\label{eq:ns1}
\e
and
\begin{eqnarray}
\left <(\Delta \hat E)^2 \right >_{\rm{HF}}\!\!\!\!\!&=&\!\!\!\!\!
\int d^2 U  \int d \nu_1 \int d \nu_2 \left(\frac{d B_{\nu_1}}{d T}\right) \left(\frac{d B_{\nu_2}}{d
  T}\right)
\nline
\!\!\!\!\!&\times&\!\!\!\!\!
\rho_N(\u,\nu_1) \rho_N(\u,\nu_2)
{S_f}^{\ast}(\u,\nu_1) {S_f}(\u,\nu_2)
\nline
\!\!\!\!\!&\times&\!\!\!\!\!
 C_{2 \pi U}(\nu_1,\nu_2)
\label{eq:fg1}
\end{eqnarray}
where  $B_{\nu}$ is the specific intensity of blackbody radiation (which can be approximated as
$2k_BT\nu^2/c^2$ in  the Rayleigh-Jeans regime) and  
$\rho_N(\u,\nu)$ is the normalized baseline distribution function
defined so that  $\int d^2U \, \int d\nu \rho_N(\u,\nu)=1$. For
a given observation,  $d^2U \, d\nu\, 
\rho_N(\u,\nu)$ is the fraction of visibilities in the interval  $d^2U \,
d\nu$ of baselines and frequency channels. Further, we expect
$\rho_N(\u,\nu) \propto \nu^{-2}$ for an
uniform distribution of the antenna separations ${\bf d}$.  

The term $\sigma$ in eq. (\ref{eq:ns1}) is the rms.  noise  expected
in an image made using  the radio-interferometric observation being
analyzed. Assuming  observations at two polarizations, we have   
\begin{equation}
\sigma =\frac{k_B T_{sys}}{  A_{eff} \sqrt{N_b t_{obs} B}}
\end{equation}
where $k_B$ is the Boltzmann constant, $T_{sys}$ the system
temperature,  $A_{eff}$ the effective collecting area of an individual
antenna in the array, $N_b$ the number of baselines, $t_{obs}$ the
total observing time and $B$ the observing  bandwidth.   

The contribution from HI fluctuations $\langle (\Delta \hat E)^2
\rangle_{\rm HF} $  is calculated using   $\left(\frac{d B_{\nu}}{d
  T}\right)$, the conversion factor  from temperature to
specific intensity at frequency $\nu$, and $C_{2 \pi   U}(\nu_1,
\nu_2)$  the multi-frequency angular power spectrum (MAPS;
\cite{kkd1}). 
The HI distribution during the epoch
of reionization is highly uncertain.  The value of $\langle (\Delta \hat E)^2
\rangle_{\rm HF}$ is sensitive to the size and clustering of the
ionized patches outside the target bubble (Paper II). Given the  lack
of information, we make the simplifying assumption  that the HI outside
the target bubble exactly traces the dark matter. This gives the most
optimistic constraints on bubble detection, the constraints are more
severe if patchy reionization is included. 

\vspace{-0.5cm}
\section{Scaling relations}
The scaling of the  expectation value of the estimator with 
various parameters can be estimated from  eq. (\ref{eq:estim2})
whereby 
\begin{equation}
\langle \hat{E} \rangle \propto  U_0^2 \, \Delta \nu_b \, \nu_c^{-2} \mid
S(0,\nu_c) \mid^2\,.
\end{equation}
Here we have  assumed that $B'$ is larger
than the frequency extent of the bubble $\Delta \nu_b$ and that the
baselines in the array  extend well beyond $U_0$. Further it is
assumed that the  array configuration  is such that the 
antenna separations ${\bf d}$ are uniformly sampled, whereby
$\rho(\u,\nu) \propto \nu^{-2}$.  Considering the noise contribution
next, it also follows from eq.(\ref{eq:ns1}) that $\langle (\Delta
\hat{E})^2 \rangle \propto 
\sigma^2 \langle \hat{E} \rangle$. We use these and the relations from
the previous Section to determine that the ${\rm SNR}$ scales as 
\begin{equation}
{\rm SNR}\propto A_{eff}\sqrt{N_b \, t_{obs}} \frac{1}{T_{sys}}
  x_{\rm HI}  \sqrt{\frac{R_b^3}{r_{\nu}^2 r'_{\nu}}}
  \frac{(1+z_c)}{H(z_c)}\,. 
\label{eq:scal1}
\end{equation}
This completely quantifies the dependence on the telescope parameters,
observation time, system temperature, neutral fraction, bubble radius
and the background expansion history.  In principle, measurements of
the SNR will provide an unique and independent way to probe the source
properties ( through $R_b$; \cite{yu05} ), inter
galactic medium, and the background
cosmology during the EoR. For the redshift range of our
interest it is reasonable to assume
$r_{\nu} \propto (1+z)^{0.25}$, $r'_{\nu} \propto (1+z)^{0.5}$ and
$H(z) \propto (1+z)^{1.5}$. Further,  for the frequency range of
our  interest the system temperature  is dominated by the sky
temperature which scales as  $T_{\rm sky}  \propto \nu^{-\beta}$ with
$\beta \sim 2.6$  which implies $T_{sys} \propto (1+z)^{\beta}$.  The
effective collecting area is nearly constant for dish antennas like
the GMRT whereas it scales as $A_{eff} \propto \nu^{-2}$ for dipoles
(eg. MWA).    Combining all of these factors we determine the scaling
of the ${\rm SNR}$ with redshift 
\begin{equation}
{\rm SNR} \propto x_{\rm HI}(z)\,  (1+z)^{\alpha}
\label{scal}
\end{equation}
where $\alpha=-\beta -1$ or $\alpha= -\beta +1$ for dish antennas or
dipoles  respectively.  
While $x_{\rm HI}$ increases with $z$, the other term
$(1+z)^{\alpha}$ has the opposite behaviour. These
two competing effects decide the redshift where  the    ${\rm SNR}$
peaks which is the  optimal redshift for bubble detection. 

The baseline distribution, in general, does not uniformly sample all
baselines. Typically, the sampling falls at larger baselines and we do
not expect the scaling relations discussed here to be exactly
valid. The deviations from the scaling relations depend on the bubble
size and the array configuration, and  in the next Section we discuss
these for the GMRT and the MWA. 

\vspace{-0.5cm}
\subsection{Evolution of neutral fraction with redshift}
\begin{figure}
\includegraphics[width=.45\textwidth]{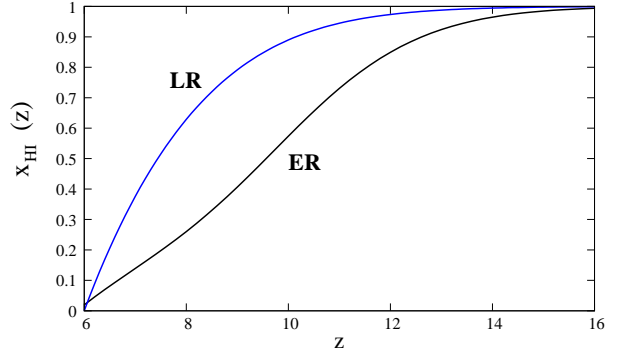}
\caption{The evolution of the mean neutral fraction
  $x_{HI}$ with redshift for the two different
  reionization models discussed in the text.}
\label{fig:zvsh1}
\end{figure}

In this work, we consider two physically motivated models of reionization,
namely, the early reionization (ER)  and the late reionization (LR) scenario.
These models are constructed using the semi-analytical formalism 
\cite{chou05,choudhury06a} which implements most of the relevant
physics governing the thermal and ionization history of the IGM, such as
the inhomogeneous IGM density distribution, three different classes of
ionizing photon sources (massive Pop III stars, Pop II stars and QSOs),
radiative feedback inhibiting star formation in low-mass galaxies and
chemical feedback for transition from Pop III to Pop II stars. The
models are consistent with various observational data, namely,
the redshift evolution of Lyman-limit absorption systems (Storrie-Lombardi
et al. 1994),
the Gunn-Peterson effect \cite{songaila}, electron scattering optical depths
\cite{kogut}, temperature of the IGM  \cite{schaye} and cosmic star formation
history \cite{nagamine}.  
We assume that these two models ``bracket'' the range of models
which are consistent with available data.

In ER scenario, hydrogen reionization starts around $z
\approx 16$ driven by metal-free (Pop III) stars, and it is $50 \%$
complete by $z \approx 10$. The contribution of Pop III stars decrease below
this redshift because of the combined action of radiative and chemical
feedback. As a result, reionization is extended considerably completing
only at $z \approx 6$ (Figure \ref{fig:zvsh1}). In LR scenario, the 
contribution from the metal-free stars
is ignored, which makes reionization start much later and is only $50 \%$
complete only around $z \approx 7.5 $. The main difference between
the ER and LR models is in their predictions for the electron
scattering optical depth (which is $0.12$ and $0.06$ for the ER
and LR scenarios, respectively).

\vspace{-0.5cm}
\section{Results and conclusions}
\begin{figure}
\includegraphics[width=.45\textwidth,angle=-90]{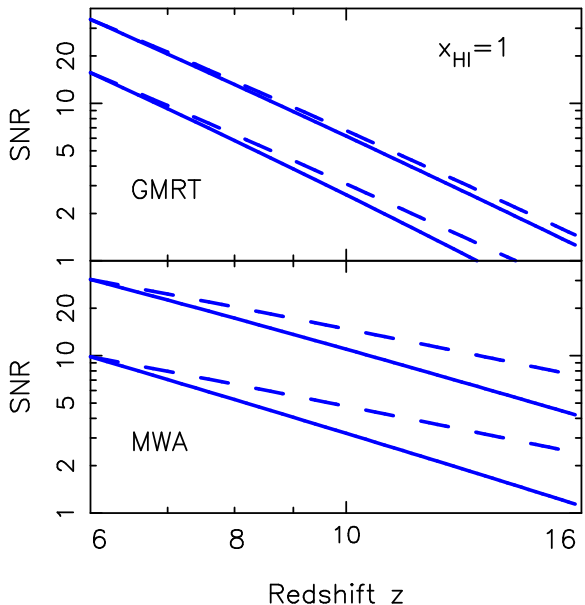}
\caption{Assuming $x_{\rm HI}=1$,
the dashed  lines show  the predicted scaling of the SNR for  uniform
baseline  
coverage (eq. \ref{scal}), the solid lines are calculated
numerically incorporating non-uniform baseline coverage.  
For both GMRT (upper panel) and MWA (lower panel), the  upper curves
are for $R_b=50 \ {\rm Mpc}$ with $1000 \ {\rm hrs}$ observation, and
the lower curves for $R_b=20 \ {\rm Mpc}$ with $4000 \ {\rm hrs}$. 
}
\label{fig:zscal}
\end{figure}

\begin{figure}
\includegraphics[width=.45\textwidth,angle=-0]{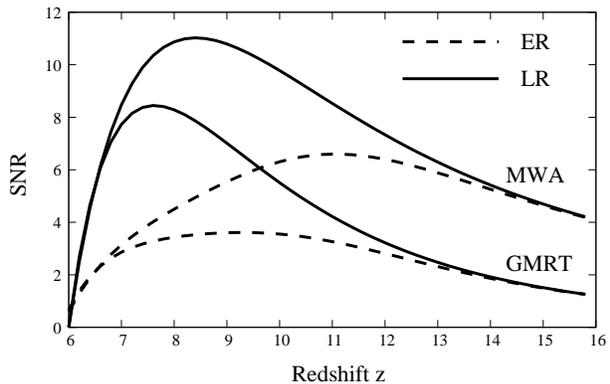}
\caption{The SNR for $R_b=50 \ {\rm Mpc}$ and $1000 \ {\rm hrs}$
  observation. Results are shown for both GMRT and MWA using the  two
different   reionization models (ER and LR)   discussed in the text.  
} 
\label{fig:zscala}
\end{figure}

\begin{figure}
\includegraphics[width=.38\textwidth, angle=270]{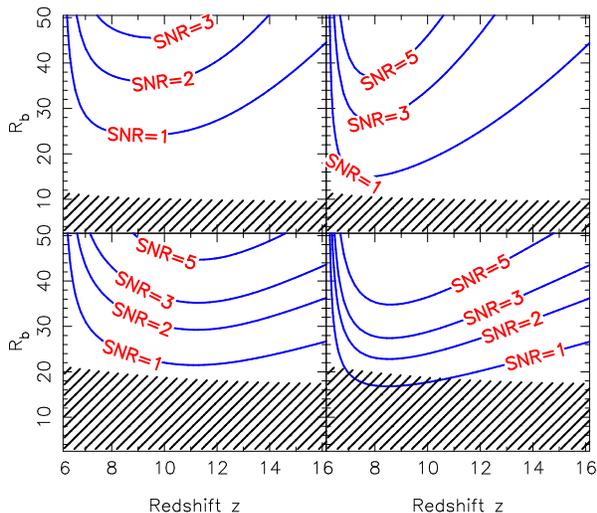}
\caption{SNR contours as a function
  of the redshift $z$ and  comoving bubble radius $R_b$,
  considering  $1000 \,
  {\rm hrs}$ of   observation with the GMRT (upper panels) and MWA
  (lower).  The left  and right panels show
 the   ER and the LR scenarios respectively. The shaded region is
  ruled out  due to  the   HI fluctuations. }  
\label{fig:z_Rf_snr}
\end{figure}

We consider two possible definitions of the 
 'optimal redshift' for bubble detection. The first is the redshift 
where, for a fixed observing time and  bubble radius $R_b$, 
the  ${\rm SNR}$ is maximum. Another possibility is, for a fixed
observing time and  ${\rm SNR}$, the redshift where a bubble of the  
smallest size can be detected.  While 
the two definitions are the same if the instrument has uniform 
baseline coverage, we do not expect this to be true in general. 

We have used equations (\ref{eq:estim2}), (\ref{eq:ns1}) and
(\ref{eq:fg1}) to calculate the ${\rm SNR}$ and determine the
constraints from HI fluctuations. The  baseline 
distribution function $\rho_N(\u,\nu)$, which we assume to be  circularly
symmetric  ($\rho_N(\u,\nu)= \rho_N(U,\nu)$), has been calculated
in Paper~I  for both the GMRT and the MWA. In both cases
$\rho_N(U,\nu)$  
falls off with increasing $U$. For the GMRT $\rho_N(U,\nu)$ is
roughly  constant for antenna separations $d < 1 \, {\rm km}$ and it extends
out to  large baselines  $d \sim 25 \, {\rm km}$. For  the MWA we have
assumed that the antennas are distributed over a circular region of
diameter $ 1.5 \, {\rm km}$, with the number density of antennas
falling as $1/r^2$ with the distance from the center. 

We first consider, for a fixed bubble
 radius  and observing time,  how the SNR varies with $z$.
Assuming  $x_{\rm HI}=1$ and uniform baseline  coverage, we expect
 that SNR$\propto 
 (1+z)^{\alpha}$ with  $\alpha=-3.6$ and  $-1.6$  for GMRT and MWA
 respectively. 
 For the GMRT,  we find (Figure  \ref{fig:zscal})
that the predicted
scaling holds for large  bubbles $R_b \geq 50 \, {\rm Mpc}$ where the
entire  signal lies within a small baseline range which is nearly
uniformly sampled.  For smaller bubbles a  significant
 amount of signal  spreads over to larger baselines which are not
 uniformly sampled.  We find that $\alpha$  changes, approximately
 linearly,  from $-3.6$  to $ -4.1$ as $R_b$ is varied from $50$ to 
 $20 \, {\rm  Mpc}$. For the MWA, the non-uniform baseline
 coverage makes the  
scaling  steeper than $-1.6$ for all values of $R_b$, and 
we find $\alpha=-2.4$ and $-2.5$ for  $R_b= 50$ and
$20 \, {\rm Mpc}$ respectively. We combine these findings  with earlier 
results at $z=8.3$  (Paper I) as to how the SNR scales with $R_b$ to
 obtain  
\begin{equation}
{\rm SNR}= x_{\rm HI} \ K \, \left( \frac{t}{1000 \ {\rm hrs}}
\right)^{0.5} \ \left( \frac{1+z}{10} \right)^{\alpha} \left(
\frac{R_b}{50 \ {\rm     Mpc}} \right) ^{\gamma} 
\label{eq:snr1}
\end{equation} 
where $t$ is the observing time, and $\alpha$, $\gamma$ and $K$ are
parameters whose values are listed in Table~\ref{tab1}. 
This expression is found to match the numerically computed SNR 
to within $20 \%$, which is quite adequate given the large uncertainty
in $x_{\rm HI}$. 

Considering  Figure \ref{fig:zscala} which shows   the SNR for  the
two reionization models , we find that it  increases monotonically
as $z$ decreases when  $x_{\rm HI}  \approx  1$ and thereafter  declines
rapidly once $x_{\rm   HI} \leq 0.5$.  The peak SNR, the corresponding
optimal 
redshift $z_o$ and the  $z$ range where the SNR is within $80 \ \%$ of
the peak value are tabulated  in Table~\ref{tab2}. Results have been
shown only  for  $R_b=50 \ {\rm Mpc}$ and $t=1000$ hrs of
observation, these can  be easily scaled to other $R_b$ and $t$ values
using eq. (\ref{eq:snr1}). The  $z$ dependence is not very different for
smaller  bubbles in the range $ 50 > R_b \ge 20 \ {\rm Mpc}$.

 The effective  collecting area of the individual MWA antennas 
 increases with  wavelength as $\lambda^{2}$. This  reduces the noise
 at higher redshifts, and puts the MWA at an advantage over the GMRT in
 detecting bubbles at high redshifts. This  also pushes the optimal
 redshift for MWA to a higher value as compared to the GMRT
 (Table~\ref{tab2}).  The MWA is also at an advantage over the GMRT in
 detecting large bubbles ($R_b \sim 50 \ {\rm Mpc}$, Figure
 \ref{fig:zscala}). The SNR scales differently with $R_b$ for the two
 instruments (Table~\ref{tab1}), and the advantage that the MWA has
 for large bubbles balances out as the bubble size is reduced.
GMRT and MWA have  nearly comparable SNR for  $R_b=30 \ {\rm  Mpc}$.

\begin{table}
\caption{Dimensionless parameters required to calculate the SNR using
  eq. (\ref{eq:snr1}). %Here  $R_b$ is in Mpc, and its values are
The values of $R_b$ are 
  restricted to the range  $50~{\rm Mpc} \ge R_b \ge  20~{\rm Mpc}$.}
\label{tab1}
\begin{tabular}{cccc}
\hline
& $K$ & $\alpha$ & $\gamma$ \\
\hline 
\hline
GMRT & $9.1$ & $-3.6 -\left(50-\f{R_b}{\rm Mpc}\right)/60$ & 1.5 \\
& & & \\
MWA & 13.4 & -2.4 & 2.0  \\
\hline
\end{tabular}
\end{table}

\begin{table}
\caption{For $R_b=50 \ {\rm Mpc}$ and $1000 \ {\rm hrs}$ of
  observation, the optimal redshift $z_o$ where
  the SNR peaks, the peak value and the $z$ range where the SNR is
  within $80 \ \%$ of the peak value.}
\label{tab2}
\begin{tabular}{ccccc}
\hline
& & $z_o$ & Peak SNR & $80 \ \%$ $z$ range\\
\hline \hline
GMRT & ER & 9.2 & 3.6 & 7 - 12 \\
  & LR & 7.6 & 8.4 & 6.8 -  9.2 \\
\hline 
MWA & ER & 11.0  & 6.59 & 8.8 - 14 \\
& LR & 8.4 & 11 & 7.1 - 10.8 \\
\hline
\end{tabular}
\end{table}

We next consider the other definition of
the optimal redshift  where  for a fixed observing time and SNR,  we
determine $z_o$ where a bubble of the smallest size can be
detected. Considering the constant SNR contours in Figure
\ref{fig:z_Rf_snr}, we  find that the $z_o$ values are roughly
consistent with those in Table~\ref{tab2}. This shows  that for
both the GMRT and the MWA, for $ 50 > R_b \ge 20 \ {\rm Mpc}$ the
two definitions predict the same optimal redshift which is
approximately independent of the bubble size. We do not expect this to
hold for smaller bubbles $R_b \sim 10 \ {\rm Mpc}$ where a detection
is possible only with the GMRT, the signal being smaller
than the HI fluctuations in  the MWA (Paper I).

Given the lack of knowledge about the reionization history, it would
be most judicious to choose a redshift where a high SNR is predicted
for both the ER and LR models. We find that the redshift range $7 -
9.2$ and $8.8 - 10.8$ are most appropriate for the GMRT and MWA
respectively. For both instruments, the prospects of a detection are
considerably improved in the late reionization scenario. Assuming
$1000 \ {\rm hrs}$  of observation,  in the ER and LR models 
respectively, a $3 \sigma$ detection is possible with the GMRT 
for $R_b \sim 50$ and $30 \ {\rm Mpc}$ or larger. The same figures   are
$40$ and $30 \ {\rm Mpc}$ for the MWA.

The actual distribution of bubble sizes is an  important issue
for bubble detection.  This depends on the reionization history and 
the distribution of ionizing sources which are largely unknown. 
 We generally expect a predominance of larger bubbles  at lower
 redshifts.  Analytic estimates \cite{furlanetto05,rhook06}  do not rule out
 bubbles in the parameter range amenable for detection with the GMRT
 and MWA.

In conclusion, we find that the optimal redshift  
for bubble detection is determined by a combination of instrument
dependent factors and the evolution of the neutral fraction $x_{\rm
  HI}$.  We propose that the redshift $8.1 \pm 1.1$ and $9.8 \pm 1$ are
optimum for detecting ionized bubbles with the GMRT and MWA
respectively. The prospects of a detection are most favourable for 
late reionization   with $x_{\rm HI} \sim 0.5$ at $z \sim 8$ where for both
instruments a $3 \sigma$ detection is possible for  $R_b \ge 30 \ {\rm 
  Mpc}$  with $1000 \ {\rm hrs}$ of observation. Future observations
will either lead to the detection of ionized bubbles, or lead to
constraints  on the product $x_{\rm HI} \ R_b^{\gamma}$ for the
observational volume in the event of non-detection. 

\vspace{-0.5cm}
\section{Acknowledgment}
KKD would like to thank CSIR, India for financial support.

\end{document}